\documentclass[preprint,aps]{revtex4}
\usepackage{graphicx}
\usepackage{dcolumn}
\usepackage{amsmath, amssymb}

\let\epsilon\varepsilon

\newcommand{\comment}[1]{}

\begin{document}

\title{Nonstationary Increments, Scaling Distributions, and Variable Diffusion Processes in Financial Markets}

\author{Kevin E. Bassler$^{1,2}$, Joseph L. McCauley$^{1,3}$, and Gemunu H. Gunaratne$^{1,4}$\footnote{Corresponding Author - Electronic Address: gemunu@uh.edu} }

\affiliation{$^1$ Department of Physics,
         University of Houston,
         Houston, TX 77204}
\affiliation{$^2$ Texas Center for Superconductivity at the University of Houston,
         Houston, TX 77204}
\affiliation{$^3$ Senior Fellow, Department of Economics,
         J.E.Cairnes Graduate School of Business and Public Policy,
         NUI Galway, Ireland}
\affiliation{$^4$ The Institute of Fundamental Studies,
         Kandy, Sri Lanka}
\maketitle

\nobreak

{\bf Arguably the most important problem in quantitative finance is 
to understand the nature of stochastic processes that underlie 
market dynamics. One aspect of the solution to this problem involves
determining characteristics of the distribution of  
fluctuations in returns. Empirical studies conducted 
over the last decade have reported that they are
non-Gaussian, scale in time, and have power-law 
(or fat) tails~\cite{mand,mccAgun,manAsta,friApei,borl}. However, 
because they use sliding interval methods of analysis, 
these studies implicitly assume that the underlying process has 
stationary increments. We explicitly show that this assumption is not valid
for the Euro-Dollar exchange rate between 1999-2004.
In addition, we find that fluctuations in returns of the exchange rate 
are uncorrelated and scale as power-laws for certain time intervals during each day.
This behavior is consistent with a diffusive process with a diffusion
coefficient that depends both on the time and the price 
change. Within scaling regions, we find that sliding interval methods 
can generate fat-tailed distributions as an artifact, and that the type 
of scaling reported in many previous studies does not exist.}

Our analysis is conducted on one-minute intra-day prices of 
the Euro-Dollar exchange rate (obtained from Olsen 
and Associates, Z\"urich) which is traded 24-hours a day.
Let $P(t)$ represent the exchange rate at time $t$ and define 
the return of the exchange rate as
${\bar x}(\tau; t) \equiv \log \left [P(\tau + t)/P(t) \right]$. 
Here $t$ represents a time during the day and $\tau$ a time increment 
that is initiated at $t$. The analysis presented below is predicated on 
the assumption, for which we provide evidence, 
that the stochastic dynamics of ${\bar x}(\tau; t)$ 
is the same between trading days. Then, we find that the average movement
taken over the approximately 1500 trading days during 1999-2004,
$\langle {\bar x} (\tau; t) \rangle$ nearly vanishes for each value of 
$t$. A value of $\tau = 10\ min$ is used so that the
autocorrelations in the signal $P(t)$ have decayed sufficiently.
The rest of our analysis is conducted on fluctuations 
$x(\tau;t) = {\bar x}(\tau;t) - \langle {\bar x} (\tau; t) \rangle$
about the mean.

A stochastic process has stationary increments if the
distribution of $x(\tau; t)$ is independent of $t$;
otherwise, increments are nonstationary.
Figure 1(a) shows the behavior of the standard deviation
$\sigma(\tau;t) \equiv \sqrt {\langle x (\tau; t)^2 \rangle}$ 
of the Euro-Dollar rate as a function
of the time of day. 
If the stochastic increments are stationary, the curve would be flat.
Clearly, it is not. Instead $\sigma (\tau; t)$ exhibits complicated
nonstationary behavior while changing by more than a factor of 
3 during the day. 

Our assumption of daily repetition of the stochastic process
is validated by conducting a corresponding analysis 
of fluctuations throughout a trading week~\cite{galAcal}. Figure 1(b) shows the 
standard deviation of returns averaged over the 300 weeks studied. The approximate 
daily periodicity of $\sigma(\tau; t)$ is evident, 
thereby justifying our approach. Similar observations were made on 
price increaments for Euro-Dollar rate in Ref.~\cite{galAcal}.

The standard deviation scales as power-laws with time during several intervals
within the day. Power-law fits to the data in some of these intervals are 
shown by colored lines in Fig.~1(a). We focus our analysis on the 
time interval {\bf I} which begins at 9:00 AM
New York time and lasts approximately $3$ hours. The data shown in red in
Fig.~2(a) shows that the standard deviation within this interval 
scales like $t^{-\eta}$ where $t$ is measured from the beginning of the 
interval and the index $\eta = 0.13 \pm 0.04$. 
This scaling extends for more than $1.5$ decades in time.
Note that the value of $\eta$ is different for the other time intervals 
during which the standard deviation scales in time. 
Similar variation in scaling exponents during the 
day has been reported previously \cite{carAcas}. 

The scaling index within {\bf I} does not change 
significantly during the six years studied. This is demonstrated by
independently analyzing three two-year periods 1999-2000, 2001-2002,
and 2003-2004. Figure 2(b) shows that the scaling index
remains nearly unchanged between these two-year periods.

We have also analyzed the behavior of other moments 
$\langle |x(\tau; t) |^{\beta}\rangle^{1/\beta}$ of the returns. 
Figure 2(a) shows that each of the moments $\beta=0.5,\ 1.0,\ 2.0$, and $3.0$
also scales as a power-law in time, and furthermore that the scaling index
for each of them is consistent with the value of $\eta = 0.15$. 
This nearly uniform scaling of the different moments suggests that the return distribution
itself scales in time. Denote the distribution of $x(\tau; t)$ by 
$W(x, \tau; t)$, where the final argument reiterates that
the distribution can depend on the starting time of the interval. In 
particular, when the increments are nonstationary $W(x,\tau; t)$ depends
on $t$. Our scaling anzatz is
\begin{equation}
  W(x, \tau; 0) = \frac{1}{\tau^{H}} {\cal F} (u)
\label{scaling}
\end {equation}
where $H$ is the {\it scaling index}, $u=x/\tau^{H}$ 
the {\it scaling variable} and ${\cal F}$ the {\it scaling function}. 
Note that the scaling anzatz is for a time interval
{\it starting} from the beginning of {\bf I}.  

In addition to scaling, the stochastic dynamics appears to have
no memory. This can be demonstrated by 
evaluating the auto-correlation function 
\begin{equation}
A_{\tau}(t_1,t_2) = \frac{\langle x(\tau;t_1) x(\tau;t_2) \rangle} {\sigma(\tau;t_1) \sigma(\tau;t_2)}.
\nonumber
\end{equation}
We find that for $\tau=10$, $A_{\tau}(t_1, t_2) = 1$ if $t_1=t_2$, and of the 
order of $10^{-3}$ when $|t_1-t_2|\ge 10$. This observation eliminates fractional 
Brownian motion~\cite{manAvan} as a description for the underlying 
stochastic dynamics, and strongly indicates that
${\partial W (x, \tau; 0)}/{\partial \tau}$ depends only on 
$x(\tau; 0)$ and $\tau$. If, in addition, $W(x, \tau;0)$ has finite 
variance (see Fig. 4), 
it has been analytically established that the 
evolution of $W(x,\tau; 0)$ is given by a diffusion equation \cite{chan,gunAmcc}
\begin {equation}
\frac {\partial W(x, \tau; 0)}{\partial \tau} = \frac{1}{2} \frac{\partial^2}{\partial x^2} \left (D(x, \tau) W(x, \tau; 0)\right),
\label{FPeqn}
\end{equation}
where $D(x, \tau)$ is the diffusion coefficient. 
There is no drift term in Eq.~(2) because 
$x(\tau; t)$ has zero mean for all $t$. 
Note that the stochastic dynamics is completely determined by the diffusion coefficient,
which, as shown below, depends on $H$. Hence, $H$ can be considered to 
be the {\it dynamical} scaling index.


Because we have found scaling, consider solutions of the form (1) to Eq.~(2). 
When $H = 1/2$, the diffusion coefficient has been 
shown to be a function of $u$; i.e., $D(x, \tau) = {\cal D}(u)$ \cite{gunAmcc}. 
If, in addition, ${\cal D}(u)$ is symmetric in $u$, it is related to the 
scaling function by 
${\cal F}(u) = D(u)^{-1} \exp \left (-\int^{u} dy \ y/D(y) \right )$ \cite{gunAmcc,aleAbas}. 
When $H \ne 1/2$, we can
``rescale" time intervals by $\tilde \tau = \tau^{2H}$~\cite{galAcal,basAgun}. In $\tilde \tau$, 
the stochastic process  has a scaling index $1/2$ and a diffusion 
coefficient of the form ${\cal D}(x/\sqrt{\tilde \tau})$. 
Converting back to $\tau$,
$D(x, \tau) = 2 H \tau^{2H-1} {\cal D}(u)$ \cite{basAgun}. 

Statistical analyses of financial markets have often been conducted using 
sliding interval methods \cite{mccAgun,manAsta,friApei,borl,borl2,galAcal,ghaAbre}, 
which implicitly assume that increments are stationary even if they are not. For example, 
they compute the distribution $W_S(x, \tau) = \langle W (x, \tau; t)\rangle_{t}$,
where $\langle . \rangle_t$ indicates an average over $t$.
Many of these studies have reported that $W_S(x, \tau)$ scales as
\begin{equation}
W_S (x, \tau) = \frac{1}{\tau^{H_S}} {\cal F}_S (v),
\label{siscaling}
\end{equation}
where $v = x / {\tau^{H_S}}$ and $H_S \approx 1/2$. 
It has also been reported that the scaling function ${\cal F}_S$  has power-law (or fat) 
tails~\cite{friApei,borl}. However, it is important to understand that $W_S(x, \tau)$ is 
a solution of Eq.~(2) only when the stochastic process has stationary
increments, in which case  $H = H_S = 1/2$. 
In general, $H_S$ and $W_S (x, \tau)$ are different from 
$H$ and $W (x, \tau; 0)$. Next, we give an explicit example where this is the case,
and, in addition, $W_S(x, \tau)$ appears to have fat-tails even though 
$W(x, \tau; 0)$ does not.

Consider a diffusive process initiated at $x=0$ that has a variable 
diffusion coefficient $2 H \tau^{2H-1} (1 + |u|)$. Its distribution
has a scaling index $H$ and a scaling function 
${\cal F}(u) = \frac{1}{2}\exp (-|u|)$~\cite{gunAmcc,aleAbas}. 
(See the discussion following Eq.~(2).) 
Numerical integration of the stochastic process for $H=0.35$ 
confirms this claim, see Fig.~3(a). 
In contrast, $W_S(x, \tau)$ calculated from {\it the same data}
appears to scale with an index $H_S = \frac{1}{2}$.
Unlike ${\cal F}$ which is bi-exponential, the apparent scaling function ${\cal F}_S$ 
(shown in Fig.~3(b)) has fat-tails.  However, a careful analysis 
reveals that distributions $W_S(x,\tau)$ do not scale in the tail region,
and hence that ${\cal F}_S$ is not well-defined.
Differences analogous to those between $H$ and $H_S$ have been
noted for L\'evy processes~\cite{fogAboh} and for the R/S analysis of Tsallis 
distributions~\cite{borl2}. 

The behavior of $\sigma(\tau; t)$ (Fig.~2(a)) can
be calculated for variable diffusion processes. Assuming that 
$\tau$ is small, Ito calculus gives 
$\delta  x^2 \equiv x(\tau; t)^2 = D(x, t) \tau$. 
Averaging over returns at $t$ gives
\begin{equation}
\langle \delta x^2 \rangle = \left[\int dx W(x, t; 0) D(x, t) \right] \tau. 
\label{mean_abs1}
\end{equation}
In a variable diffusion process,
$W(x,t; 0) = t^{-H} {\cal F}(u)$ and $D(x; t) = 2 H t^{2H -1} {\cal D}(u)$;
consequently
\begin{equation}
\sqrt {\langle \delta x^2 \rangle} \sim t^{H - 1/2},
\label {mean_abs2}
\end{equation}
independent of the exact form of ${\cal D}(u)$.
Results for the Euro-Dollar rate within the interval {\bf I} (Fig.~2(a)) 
which showed that $\eta \approx 0.15$ are therefore consistent with a 
scaling index $H = \frac{1}{2} - \eta \approx 0.35$. Note that, unlike for
L\'evy processes and fractional Brownian motion, $H < 1/2$, and is substantially less than $H_S$
reported in previous analyses of the Euro-Dollar exchange rate 
(between 0.5 and 0.6)~\cite{galAcal,ghaAbre,mulAdac}.
A general calculation for the moments of a variable diffusion process gives
\begin{equation}
\langle |\delta x|^{\beta} \rangle^{1/\beta} \sim t^{H -1/2},
\label {mean_abs3}
\end{equation}
for all $\beta$, consistent with results shown in Fig.~2(a).

In order to estimate $H_S$ for an arbitrary variable diffusion process, we note first that 
$\langle x(t+\tau;0)^2\rangle = \langle x(t;0)^2 \rangle + \langle x(\tau;t)^2 \rangle$
for any diffusive process without memory (see Ref.\cite{gunAmcc}). 
Then, using the scaling anzatz (1), setting $c = \int du \ u {\cal F}(u)$,
and taking the sliding interval average
\begin{equation}
\langle x(\tau; t)^2 \rangle_t = \langle c (t+\tau)^{2H} - c t^{2H}\rangle_t \approx 2Hc \langle t^{2H-1}\rangle_t \tau,
\label{expn}
\end{equation}
where the last approximation is valid when $\tau \ll t$, a condition
that is true for most intervals of 
length $\tau$ in a sliding interval calculation. Hence 
$\langle x(\tau; t)^2 \rangle_{t} \sim \tau$. 
Consequently, $H_S = 1/2$ regardless of the value of $H$!

Finally, we introduce a method to extract the empirical scaling function ${\cal F}$ from 
the Euro-Dollar time series. Unfortunately, the available data
are insufficient to determine ${\cal F}(u)$ accurately using the usual method
of collapsing $W(x, \tau; 0)$ for multiple values of $\tau$. However, since we have 
determined $H (\approx 0.35)$ independently, we can use Eq.~(\ref{scaling}) for 
multiple values of $\tau$ in the interval {\bf I} (i.e., $\tau$ 
between approximately $10$ and $160$ minutes) to determine ${\cal F}$.
The result is shown in Fig.~4(a). Note that the distribution has an approximate bi-exponential 
form. Since exponential distributions have finite variance, all assumptions
needed for the derivation of Eq.~(\ref{FPeqn}) are justified.
However, it is asymmetric and decays more slowly on the negative side.
By contrast, the empirical sliding interval scaling function ${\cal F}_S(v)$ 
for the same time interval is shown in Fig.~4(b). 
For this case, the scaling collapse is achieved for $H_S=1/2$. $F_S(v)$ appears to have fat tails, 
consistent with previous reports~\cite{mulAdac,borl}. 
However, in light of the example discussed earlier and the fact that
$H \ne 1/2$, it is unlikely that ${\cal F}_S$ is well-defined
for this financial market data within the interval {\bf I}.

Variable diffusion processes exhibit another signature (stylized
fact) of market fluctuations. Although their 
autocorrelation vanishes, a large fluctuation will typically produce 
a large value of $|x|$, and hence a return with 
a large diffusion coefficient. Consequently, a large 
fluctuation is likely to be followed by additional large fluctuations
whose signs are uncorrelated to the first~\cite{gunAmcc}. As a result, the 
autocorrelation function for the signal $|x(\tau;t)|$ (or for the signal $x(\tau;t)^2$) 
will decay slowly in $t$. Such behavior, 
referred to as the ``clustering of volatility" is seen in the 
Euro-Dollar exchange rate and has been reported in 
empirical studies of other financial markets~\cite{conApot,heyAyan,heyAleo}.

The analysis given here applies to stochastic dynamics of a single
scaling interval. However, the daily fluctuations in the 
Euro-Dollar rate are a combination of scaling intervals with distinct
scaling indices, and possibly regions with no scaling.
We have not yet determined how to extend our analysis 
beyond a single scaling region. Bacuase of this, it is not clear how 
to interpret the distributions over intervals longer than 
a scaling region, including inter-day data. 

We have shown that stochastic fluctuations in the Euro-Dollar 
rate have uncorrelated nonstationary increments during the course of a trading day,
and that there are intervals during which their absolute moments scale like 
a power-law in time. The stochastic dynamics during these scaling intervals
can be described by a diffusion process with variable diffusion coefficient.
We have also shown that sliding interval analysis of variable diffusion processes can give an 
incorrect scaling exponent and in addition can give scaling functions with
fat-tails even when the underlying dynamics do not have them. Indeed, this
appears to be the case within the interval {\bf I}.


The authors would like to thank A. A. Alejandro-Quinones for discussions.
They also acknowledge support from the Institute for Space Science
Operations (KEB, GHG) and the NSF through grants DMR-0406323 (KEB), 
DMR-0427938 (KEB) and DMS-0607345 (GHG).

\begin{figure}
\includegraphics[width=2.5in]{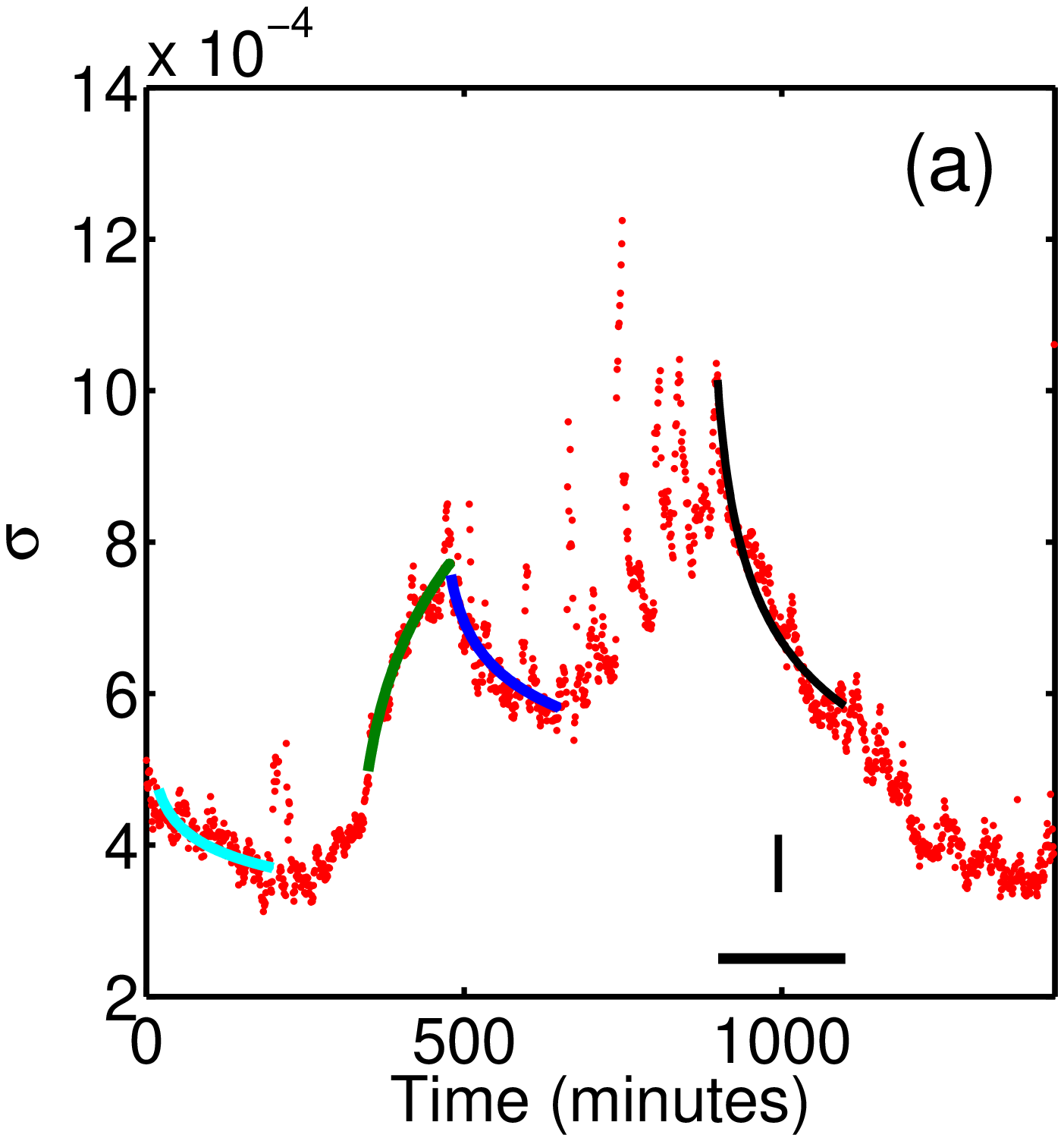}\\
\includegraphics[width=2.5in]{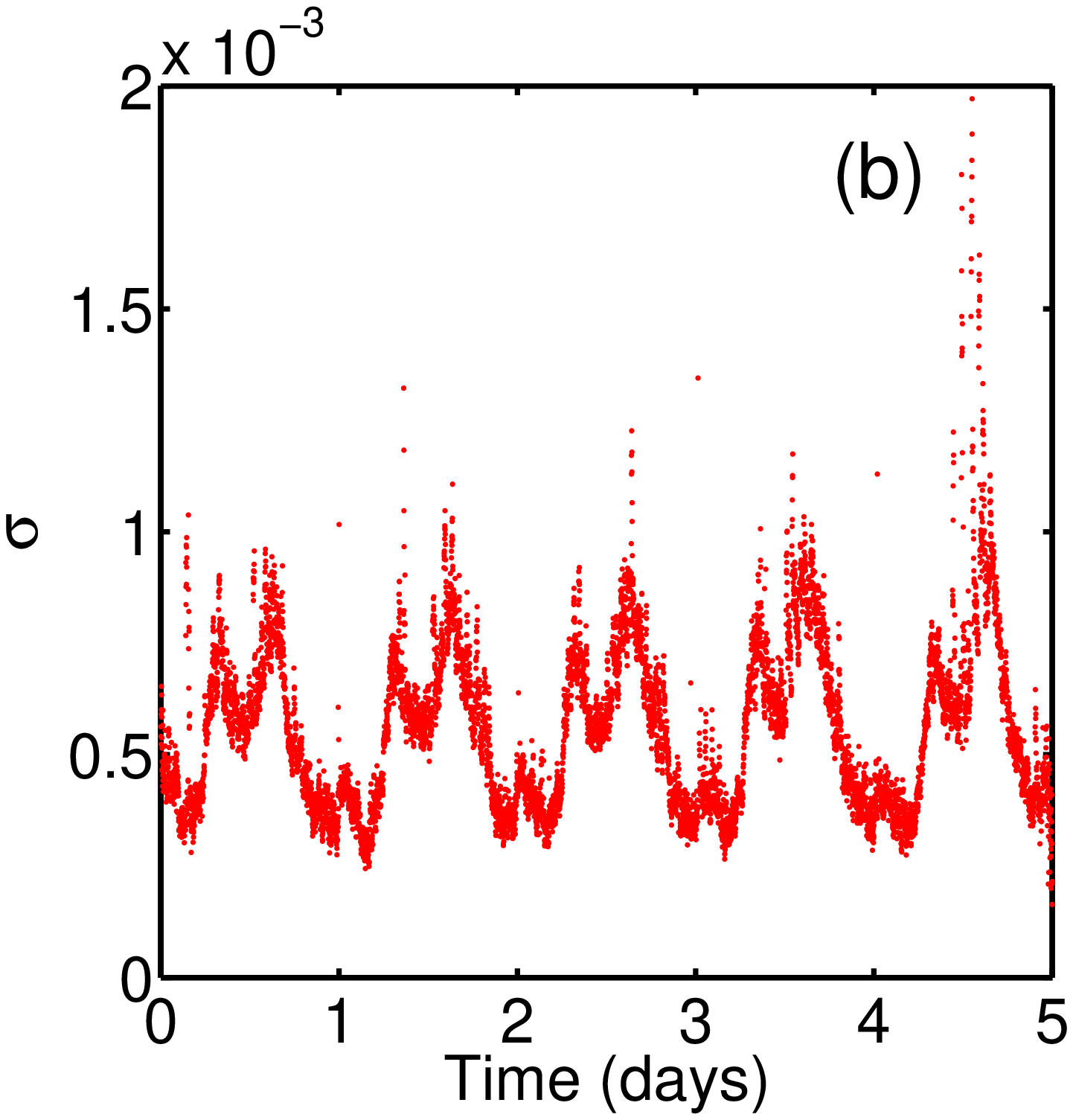}
\caption {(a) The standard deviation 
$\sigma(\tau;t) \equiv \sqrt {\langle x (\tau; t)^2 \rangle}$
of the daily Euro-Dollar exchange as a function of the time of day (in GMT).
Here $\tau = 10\ min$ to ensure that autocorrelations in $P(t)$ have decayed sufficiently.
Our statistical analysis assumes that $x (\tau; t)$ follows the same stochastic 
process each trading day. The average indicated by 
the brackets $\langle . \rangle$ is taken over the 
approximately 1500 trading days between 1999-2004, and the standard error
at each point is typically 3\%. Note that, if the stochastic dynamics had stationary
increments, $\sigma(\tau;t)$ would be constant. Instead, it varies by more 
than a factor of 3 during the day, thus 
showing explicitly that the exchange rate has nonstationary 
increments. Notice also that $\sigma(\tau;t)$ scales in time during several
intervals, four of which are highlighted by colored lines that are power-law fits. 
Our analysis focuses on the interval {\bf I} shown by the horizontal solid line.
(b) The weekly behavior of $\sigma(\tau;t)$ for the same data. Observe that it exhibits an
approximate daily periodicity, thereby justifying our assumption of the daily repeatability
of the stochastic process underlying the Euro-Dollar exchange rate.}
\label{figure1}
\end{figure}

\begin{figure}
\includegraphics[width=2.5in]{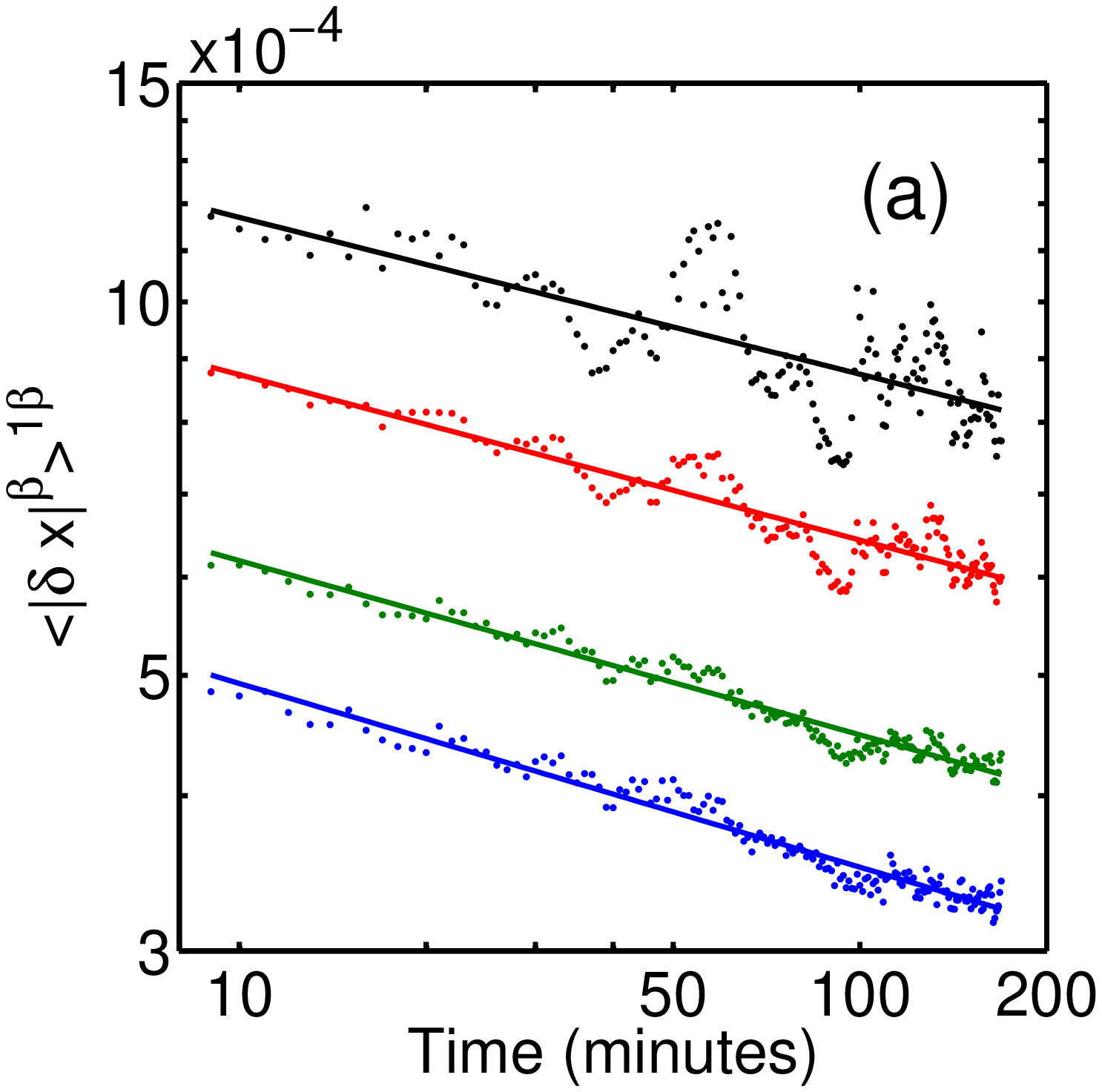}\\
\includegraphics[width=2.5in]{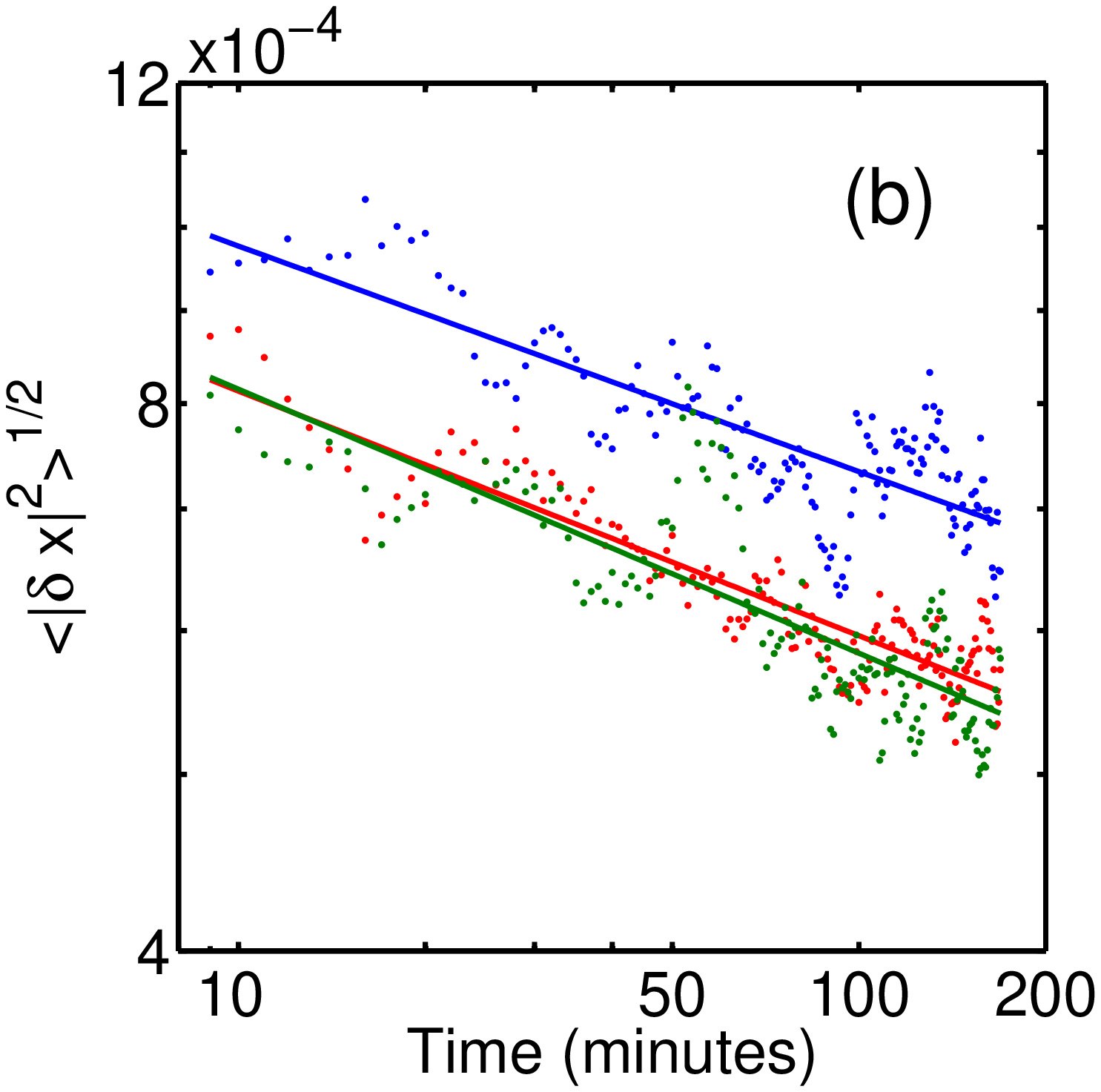}
\caption {(a) A log-log plot of $\langle x(\tau; t)^{\beta} \rangle^{1/\beta}$ 
for $\beta = 0.5, 1.0, 2.0$, and $3.0$, demonstrating power law decay $t^{-\eta}$ for each index.
Here $t$ is measured in local New York time stating at 9:00AM. 
The data for $\beta = 0.5, 1.0, 2.0$, and $3.0$, shown in blue, green, red, and 
black, respectively, have scaling indices (given by the slopes of the solid lines) 
$\eta = 0.15\pm0.02$, $0.14\pm0.02$, $0.13\pm0.04$ and $0.13\pm0.08$.
All of these values are consistent with $\eta \approx 0.15$, and hence
a dynamical scaling index of $H=\frac{1}{2}-\eta \approx 0.35$. The error estimates on the exponents 
are the standard errors from the nonlinear fit including the standard 
deviations for each time point, but neglecting any correlations between them.
(b) The behavior of the standard deviation $\sigma(\tau;t)$
in the interval {\bf I} during each of the periods 1999-2000 (blue), 2001-2002 (red), and 
2003-2004 (green). The scaling index from nonlinear fits for the three data sets are 
$0.13 \pm 0.06$, $0.14 \pm 0.04$ and $0.14 \pm 0.07$.
The near equality of these indices shows that the scaling index is nearly 
invariant over time. }
\label{Figure2}
\end{figure}

\begin{figure}
\includegraphics[width=2.5in]{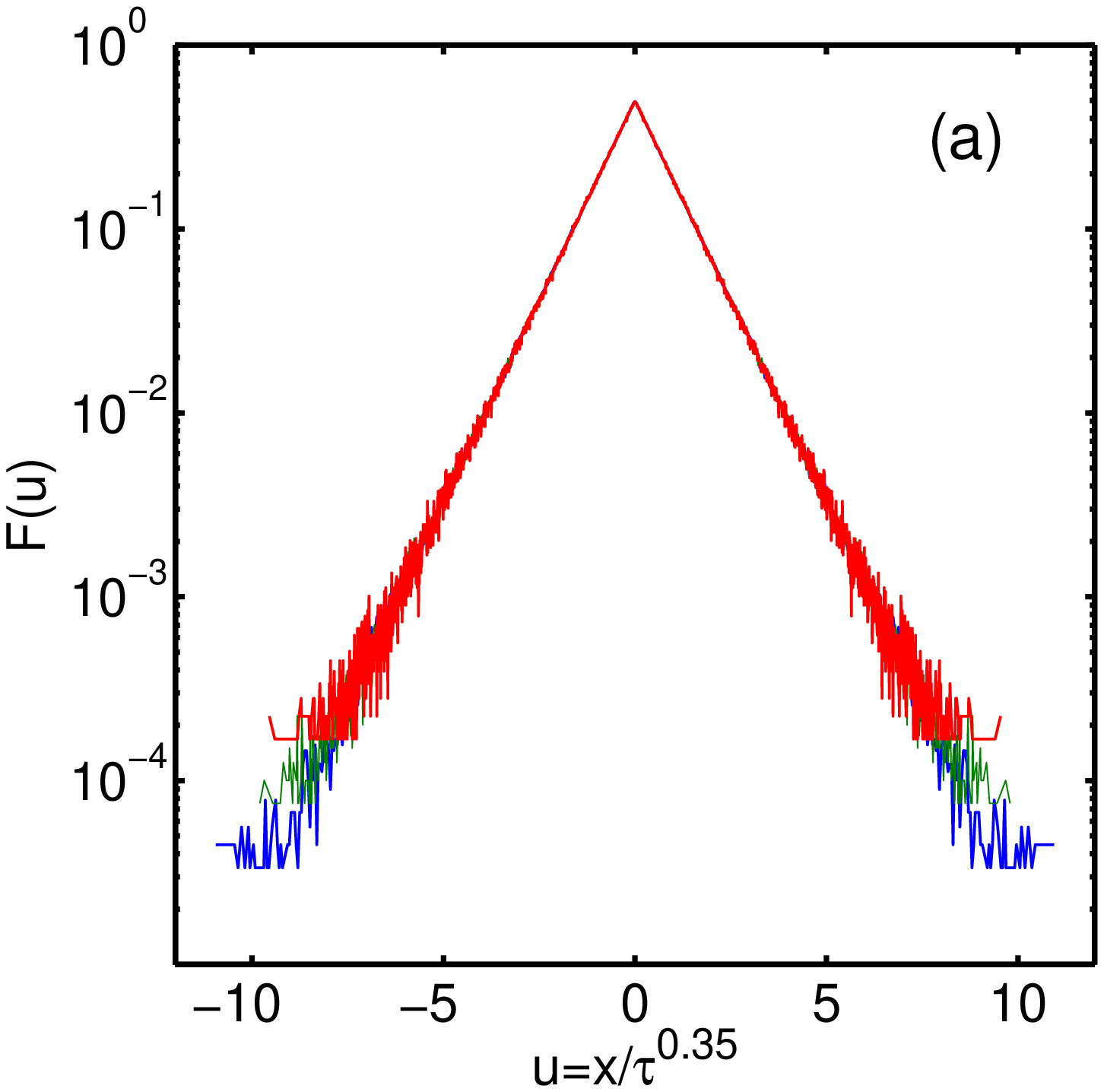}\\ ~ \\
\includegraphics[width=2.5in]{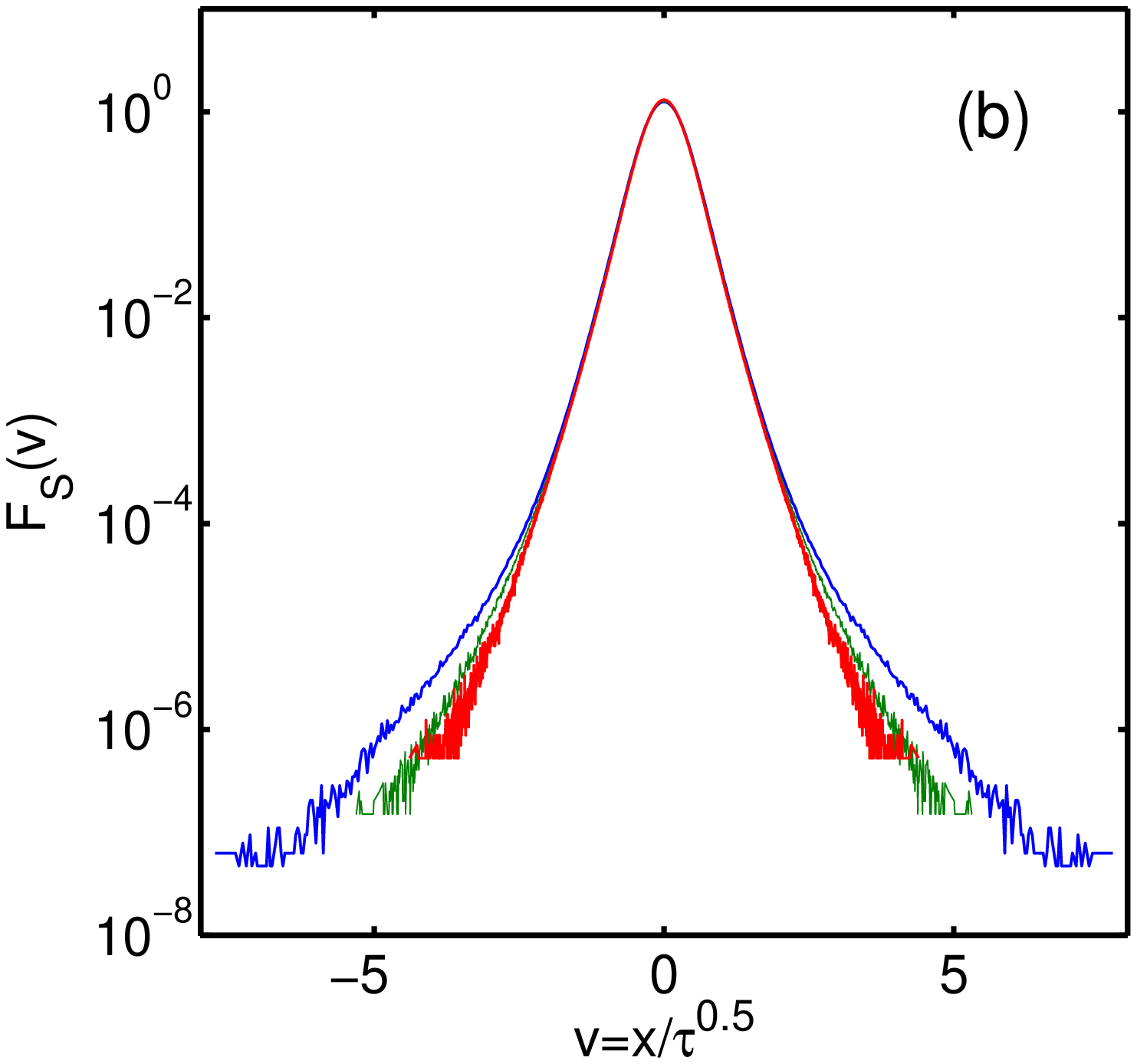}
\caption {(a) The scaling function of the return distribution ${\cal F}$ calculated 
from a collapse of data for $\tau=10$ (blue), $100$ (green), and $1000$ (red) units.
The results are from a set of 5,000,000 independent stochastic processes with
variable diffusion. The scaling index used was $H=0.35$,
and the diffusion coefficient was $2H t^{2H-1} (1+|u|)$.
Note that ${\cal F}$ is bi-exponential, as discussed in the text. 
(b) The sliding interval scaling function ${\cal F}_S$ calculated from the same 
runs. Shown are results for sliding intervals with $\tau=10$ (blue), $100$ (green) and 
$1000$ (red) units from runs of length $10,000$ units. Unlike ${\cal F}$, it appears to have fat tails.  
The scaling index used here for which the scaling collapse is achieved is $H_S =  1/2$ 
even though the dynamical scaling index 
is $H = 0.35$. Note, however, although the central part of 
the distribution scales well, the tails do not.}
\label{Figure3}
\end{figure}

\begin{figure}
\includegraphics[width=2.5in]{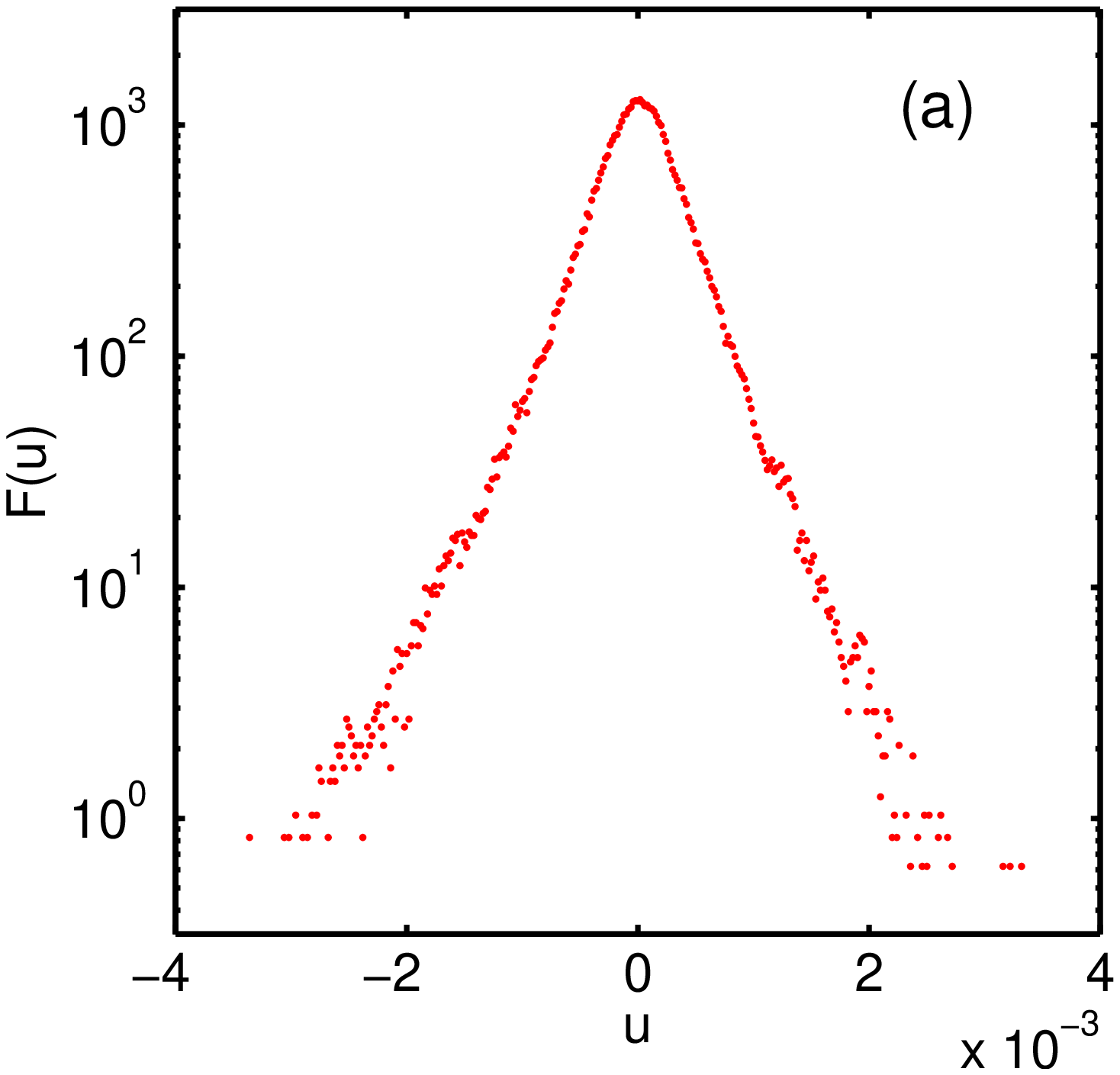}\\ ~ \\
\includegraphics[width=2.5in]{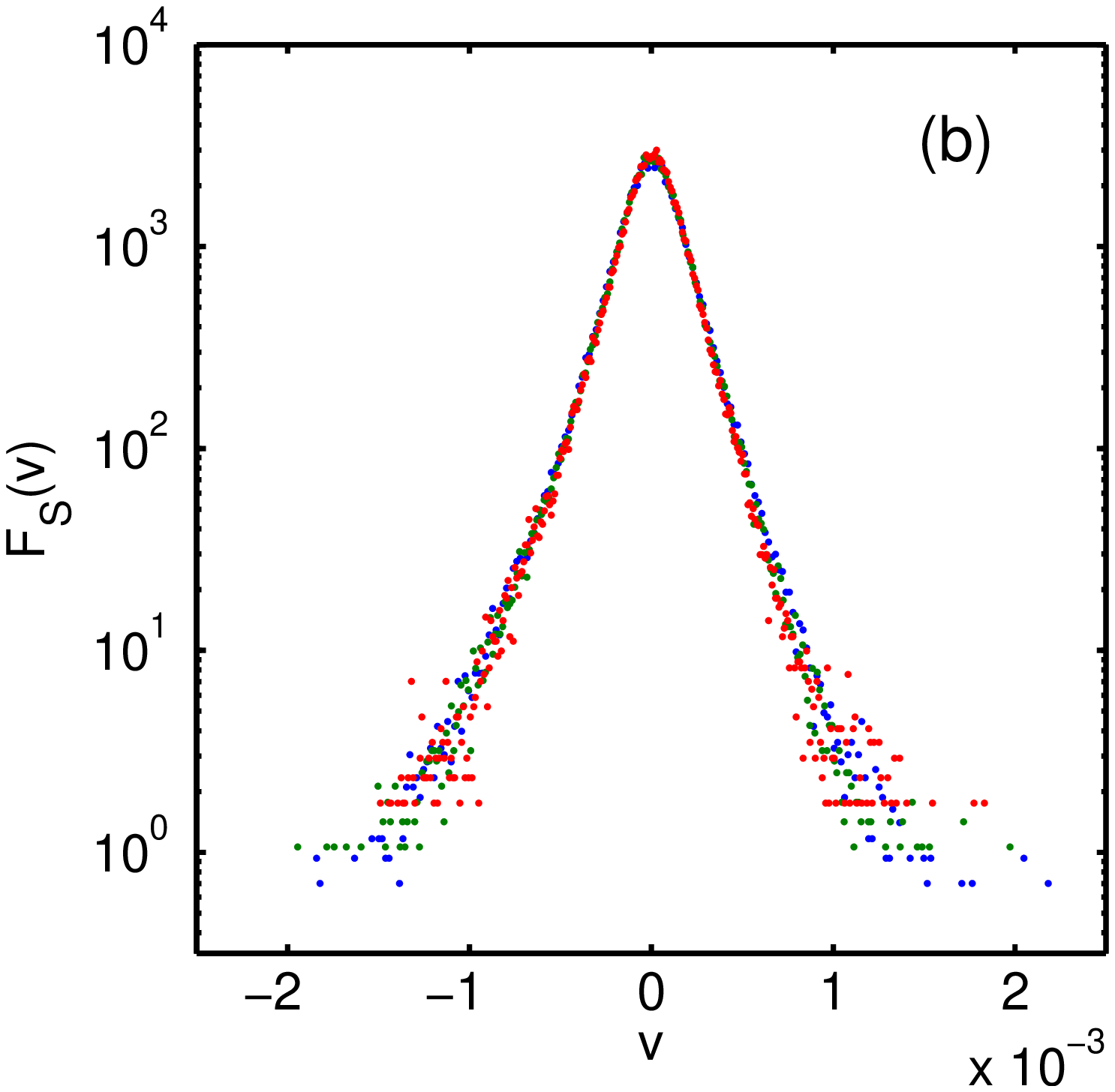}
\caption {(a) The empirical scaling function ${\cal F}$ for interval {\bf I} calculated assuming 
the scaling anzatz Eq.~(1) with $H=0.35$ and values of $\tau$ between
10 and 160 minutes. Note that ${\cal F}$ is slightly asymmetric and approximately 
bi-exponential. Since exponential distributions have finite variance,
all assumptions needed for the derivation of Eq.~(\ref{FPeqn}) are 
justified. (b) The empirical sliding interval scaling 
function ${\cal F}_S$ for interval {\bf I}
calculated by scaling collapse of data using the anzatz Eq.~(3) for 
$\tau$ of 10 (blue), 20 (green) and 40 (red) minutes. 
Note that ${\cal F}_S$ has fat-tails.}
\label{Figure4}
\end{figure}


\begin{thebibliography}{99}

\bibitem{mand} B. B. Mandelbrot, The Variation of Certain Speculative Prices, 
  {\em J. Bus.} {\bf 36}, 394 (1963).

\bibitem{mccAgun} J. L. McCauley and G. H. Gunaratne, 
  An Empirical Model of Volatility Returns and Options Pricing,
  {\em Physica A} {\bf 329}, 170 (2003).

\bibitem{manAsta} R. N. Mantegna and H. E. Stanley, 
  Scaling Behavior in the Dynamics of an Economic Index, {\em Nature} {\bf 376}, 46 (1995);
  Turbulence in Financial Markets, {\em Nature} {\bf 383}, 587 (1996).

\bibitem{friApei} R. Friedrich, J. Peinke, and Ch. Renner, 
  How to Quantify Deterministic and Random Influences on the Statistics 
  of the Foreign Exchange Market,
  {\em Phys. Rev. Lett.} {\bf 84}, 5224 (2000).

\bibitem{borl} L. Borland, A Theory of Non-Gaussian Option Pricing,
  {\em Quan. Finance} {\bf 2}, 415 (2002).

\bibitem{galAcal} S. Galluccio, G. Caldarelli, M. Marsili, and Y. C. Zhang, 
  Scaling in Currency Exchange,
  {\em Physica A} {\bf 245}, 423 (1997).

\bibitem{carAcas} A. Carbone, G. Castelli, and H. E. Stanley, 
  Time-dependent Hurst Exponents in Financial Time Series, 
  {\em Physica A} {\bf 344}, 267 (2004).

\bibitem{manAvan} B. Mandlebrot and J. W. van Ness,
  Fractional Brownian Motion, Fractional Noise and Applications,
  {\em SIAM Rev.} {\bf 10}, 422 (1968).

\bibitem{gunAmcc} G. H. Gunaratne, J. L. McCauley, M. Nicole, and A. T\"or\"ok.
  Variable Step Random Walks and Self-Similar Distributions,
  {\em J. Stat. Phys.} {\bf 121}, 887 (2005).

\bibitem{chan} S. Chandrasekhar, 
  Stochastic Problems in Physics and Astronomy,
  {\em Rev. Mod. Phys.}, {\bf 15}, 1 (1943).


\bibitem{aleAbas} A. A. Alejandro-Quinones, K. E. Bassler, M. Field, J. L. McCauley, M. Nicol,
  I. Timofeyev, A. T\"or\"ok, and G. H. Gunaratne, 
  A Theory of Fluctuations in Stock Prices,
  {\em Physica A} {\bf 363}, 383 (2006).

\bibitem{basAgun} K. E. Bassler, G. H. Gunaratne, and J. L. McCauley, 
   Markov Processes, Hurst Exponents, and Nonlinear Diffusion Equations with
   Applications to Finance, Physica A, {\bf 369}, 343 (2006).

\bibitem{borl2} L. Borland, 
  Microscopic Dynamics of the Nonlinear Fokker-Planck Equation: A Phenomenological Model,
  {\em Phys. Rev. E} {\bf 57}, 6634 (1998).

\bibitem{ghaAbre} S. Ghashghaie, W. Breymann, J. Peinke, P. Talkner, and Y. Dodge,
  Turbulent Cascades in Foreign Exchange Markets,
  {\em Nature} {\bf 381}, 767 (1996).

\bibitem{fogAboh} H. C. Fogedby, T. Bohr, and H. J. Jensen, 
  Fluctuations in a L\'evy Flight Gas,
  {\em J. Stat. Phys.} {\bf 66}, 583 (1992).

\bibitem{mulAdac} U. A. M\"uller, M. M. Dacorogna, R. B. Olsen, O. V. Pictet, M. Schwarz, and C. Morgenegg,
  Statistical Study of Foreign Exchange Rates, Empirical Evidence of a Price Change Scaling Law, 
  and Inter-day Analysis,
  {\em J. Bank. Fin.} {\bf 14}, 1189 (1990).

\bibitem{conApot} R. Cont, M. Potters, and J.-P. Bouchard, 
  Scaling in Stock Market Data: Stable Laws and Beyond,
  in "Scale Invariance and Beyond", eds. B. Dubrulle, F. Graner, and D. Sornette,
  Springer, Berlin, 1997.

\bibitem{heyAyan} C. C. Heyde and Y. Yang, 
  On Defining Long Range Dependence,
  {\em J. Appl. Prob.} {\bf 34}, 939 (1997).

\bibitem{heyAleo} C. C. Heyde and N. N. Leonenko, 
  Student Processes,
  {\em Adv. Appl. Prob.} {\bf 37}, 342 (2005).

\bibitem{couAdav} M. Couillard and M. Davison, A Comment on Measuring the Hurst
  exponent of Financial Time Series,
  {\em Physica A} {\bf 348}, 404 (2005).

\end{thebibliography}
\end{document}